\documentclass[preprint,pre,nobibnotes,twocolumn,10pt]{revtex4}%
\usepackage{amsfonts}
\usepackage{amsmath}
\usepackage{amssymb}
\usepackage{graphicx}%
\providecommand{\U}[1]{\protect\rule{.1in}{.1in}}

\begin{document}
\preprint{ }
\preprint{UATP/1502}
\title{Determination of Nonequilibrium Temperature and Pressure using Clausius
Equality in a State with Memory: A Simple Model Calculation}
\author{P.D. Gujrati,$^{1,2}$ Iakov Boyko$^{1,2}$ and Tyler Johnson$^{1,3}$}
\affiliation{$^{1}$Department of Physics, $^{2}$Department of Polymer Science, $^{3}%
$Department of Mechanical Engineering, The University of Akron, Akron, OH 44325}
\email{pdg@uakron.edu}

\begin{abstract}
Use of the extended definition of heat $dQ=d_{\text{e}}Q+d_{\text{i}}Q$
converts the Clausius inequality $dS\geq d_{\text{e}}Q/T_{0}$ into the
Clausius equality $dS\equiv dQ/T$ involving the nonequilibrium temperature $T$
of the system having the conventional interpretation that heat flows from hot
to cold. The equality is\ applied to the exact nonequilibrium quantum
evolution of a $1$-dimensional ideal gas free expansion. In a first ever
calculation of its kind in an expansion which retains the memory of initial
state, we determine the nonequilibrium temperature $T$ and pressure $P$, which
are then compared with the ratio $P/T$ obtained by an independent method to
show the consistency of the nonequilibrium formulation. We find that the
quantum evolution by itself cannot eliminate the memory effect; hence, it
cannot thermalize the system.

\end{abstract}
\date{\today}
\maketitle

There seems to be a lot of confusion about the meaning of temperature,
pressure, etc. in nonequilibrium thermodynamics \cite[for example]%
{Muschik,Keizer,Morriss,Jou,Hoover,Ruelle}, where different definitions lead
to different results. In contrast, the meaning of temperature in equilibrium
thermodynamics as $T=dQ/dS$ has no such problem, even though Planck
\cite{Planck} had already suggested that it should be defined for
nonequilibrium states just as entropy is defined. The temperature was
apparently first introduced by Landau \cite{Landau0} for partial set of the
degrees of freedom. Consider a system $\Sigma$ (in a medium $\widetilde{\Sigma
}$, which is\emph{ always} taken to be in equilibrium at temperature $T_{0}$,
pressure $P_{0}$, etc.) that was initially in an equilibrium state
A$_{\text{i,eq}}$; its equilibrium entropy $S_{\text{i,eq}}(T_{0}$,$P_{0})$
can also be written as $S_{\text{i,eq}}(E_{\text{i}}$,$V_{\text{i}})$, where
$E_{\text{i}}$,$V_{\text{i}}$ are the energy and volume of the system and the
suffix i denotes the initial state. If $\Sigma$ is now isolated from
$\widetilde{\Sigma}$, it will remain in equilibrium \emph{forever} unless it
is disturbed and all its properties such as its temperature, pressure, energy,
etc. are well defined and time invariant. Let us now disturb $\Sigma$ at time
$t=0$ by bringing it in \emph{athermal} contact (no heat exchange) with some
working medium $\widetilde{\Sigma}^{\prime}$ at pressure $P_{0}^{\prime}\neq
P_{0}$, etc. We can also disturb $\Sigma$ at time $t=0$ by bringing it in
\emph{thermal} contact (resulting in heat exchange but no work exchange) with
some thermal medium $\widetilde{\Sigma}^{\prime\prime}$ at temperature
$T_{0}^{\prime\prime}\neq T_{0}$. As $\Sigma$ tries to come to equilibrium, we
can ask: what are $\Sigma$'s temperature $T(t)$, pressure $P(t)$, etc.,
examples of its instantaneous fields, if they can be defined \emph{during}
these nonequilibrium processes? To be consistent with the second law, we need
to ensure that the definition of instantaneous pressure and temperature must
result in irreversible work that is always \emph{nonnegative}, and that heat
always flows from hot to cold. To the best of our knowledge, this question has
not been answered satisfactorily
\cite{Muschik,Keizer,Morriss,Jou,Hoover,Ruelle} for an arbitrary
nonequilibrium state. The question is not purely academic as it arises in
various contexts of current interest in applying nonequilibrium thermodynamics
to various fields such as the Szilard engine \cite{Marathe,Zurek,Kim},
stochastic thermodynamics \cite{Siefert}, Maxwell's demon
\cite{Wiener,Brillouin}, thermogalvanic cells, corrosion, chemical reactions,
biological systems \cite{Hunt,Horn,Forland}, etc. to name a few.

\ \textsc{background }Recently, we have proposed
\cite{Gujrati-Heat-Work,Gujrati-I,Gujrati-II,Gujrati-III,Gujrati-Symmetry} a
definition of the nonequilibrium temperature, pressure, etc. for a
nonequilibrium system that is in internal equilibrium; the latter requires
introducing \emph{internal variables} $\boldsymbol{\xi}$ as additional
\emph{state variables} that become superfluous in the equilibrium state. Here,
we extend the definition of these fields for $\Sigma$ in any arbitrary state
and verify its consistency with the second law by providing an alternative but
physically more intuitive approach. The entropy $S$ in an arbitrary state may
have a memory of the initial state so that it is \emph{not} a state function.
Such a memory is encoded in the probabilities $\left\{  p_{k}(t)\right\}  $,
$k$ denoting $\Sigma$'microstates, and is absent for a system in equilibrium
or in internal equilibrium for which $S$ is a state function. In terms of
$\left\{  p_{k}(t)\right\}  $ and energies $\left\{  E_{k}(t)\right\}  $, the
entropy and energy are given as $S(t)=-%
{\textstyle\sum\nolimits_{k}}
p_{k}\ln p_{k}$ and $E(t)=%
{\textstyle\sum\nolimits_{k}}
E_{k}p_{k}$, respectively, even if $S$ is not a state function
\cite{Gujrati-Entropy1,Gujrati-Entropy2}. We can identify the two
contributions in the first law $dE(t)=dQ(t)-dW(t)$
\cite{Gujrati-Heat-Work,Gujrati-Entropy1,Gujrati-Entropy2} for any
\emph{arbitrary} infinitesimal process as%
\begin{equation}
dW\equiv-%
{\textstyle\sum\nolimits_{k}}
p_{k}dE_{k},dQ\equiv%
{\textstyle\sum\nolimits_{k}}
E_{k}dp_{k}. \label{dQ-dW}%
\end{equation}
The microstate representation ensures that both $dW$ and $dQ$\ are defined for
any arbitrary process in terms of changes $\left\{  dE_{k}\right\}  $ and
$\left\{  dp_{k}\right\}  $; in addition, they depend only on the quantities
pertaining to the system \cite{Gujrati-Heat-Work,Gujrati-I,Gujrati-Entropy2}
and not those of the medium. This makes dealing with system's properties
extremely convenient. As $dW(t)$ contains \emph{fixed} $p_{k}$s so that $S$
remains fixed, it represents an \emph{isentropic} quantity to be identified as
work \cite{Landau}. As $dQ(t)$ contains the changes $dp_{k}$s, which also
determine the entropy change $dS(t)=-%
{\textstyle\sum\nolimits_{k}}
dp_{k}\ln p_{k}$, the two quantities must be related. In the following, we
only consider a \emph{macroscopic} system. Assuming both quantities to be
\emph{extensive}, this relationship must be \emph{always} linear, resulting in
the\emph{ Clausius equality }\cite{Gujrati-I,Gujrati-II,Gujrati-Entropy2}:%
\begin{equation}
dQ(t)\equiv T(t)dS(t), \label{dQ-dS}%
\end{equation}
with the intensive field $T(t)$ identified as the \emph{statistical}
definition of the temperature of $\Sigma$ so that heat flows from hot to cold
as shown below. We only consider \emph{positive} temperatures here. It may
have a complicated dependence on state variables and memory through the
dependence of $\left\{  p_{k}(t)\right\}  $ on the history. The work as a
statistical average of $-dE_{k}$ remains true in general for all kinds of work
including those due to $\boldsymbol{\xi}$. If $dE_{k}$ is only due to volume
change $dV$, then $dW(t)=P(t)dV$, which is also linear in $dV(t)$ as assumed
above; here $P(t)\equiv-%
{\textstyle\sum\nolimits_{k}}
p_{k}P_{k}$ is the average pressure on the walls (during any arbitrary
process) with a similar complicated dependence through $\left\{
p_{k}(t)\right\}  $, and $P_{k}\equiv-\partial E_{k}/\partial V$ is the
outward pressure, independent of the process, that is exerted by the $k$th
microstate \cite{Landau-QM}. It immediately follows in this case that
$dQ(t)\equiv\left.  dE(t)\right\vert _{V}$ so that the statistical temperature
is \emph{also} the thermodynamic temperature $\partial E/\partial S$. It can
be shown that in general, $T(t)$ and $\partial E/\partial S$ are the same for
a system in internal equilibrium \cite{Gujrati-I,Gujrati-II,Gujrati-Entropy2}
so that the $t$-dependence in $T(t)$ is due to the $t$-dependence of the state
variables. This makes $T(t)$ a \emph{state function}. It is no longer a state
function for a state with memory. Same comments apply to $P(t)$ or other fields.

It should be clear that $\Sigma$'s internal pressure $P(t)$, etc. have no
relationship with the \emph{external} pressure $P_{0}^{\prime}$, etc. (except
in equilibrium). Thus, $dW(t)$ is in general not the negative of the work done
$d\widetilde{W}(t)$ $[\equiv-d_{\text{e}}W(t)]$ by $\widetilde{\Sigma}$
\cite{deGroot,Prigogine,note-0} on $\Sigma$. The net work $d_{\text{i}%
}W(t)\equiv dW(t)$ $+d\widetilde{W}(t)\equiv dW(t)$ $-d_{\text{e}}W(t)\geq0$
is irreversibly dissipated in the form of heat $d_{\text{i}}Q(t)$
\cite{note-0} generated within the system; see below. It follows then that
$dQ(t)$ cannot represent the exchange heat $d_{\text{e}}Q(t)=T_{0}d_{\text{e}%
}S(t)\leq T_{0}dS(t)$ (\emph{Clausius inequality}) between $\Sigma$ and
$\widetilde{\Sigma}$. To fully appreciate this point, we recognize that the
change $dp_{k}(t)\equiv d_{\text{e}}p_{k}(t)+d_{\text{i}}p_{k}(t)$
\cite{note-0} consists of two parts: the change $d_{\text{e}}p_{k}$ caused by
the interaction of the system with the medium\ and $d_{\text{i}}p_{k}$ by the
irreversible processes going on inside the system. Accordingly, $dQ\equiv
d_{\text{e}}Q+d_{\text{i}}Q$ with $d_{\text{e}}Q=%
{\textstyle\sum\nolimits_{k}}
E_{k}d_{\text{e}}p_{k}$ and $d_{\text{i}}Q=%
{\textstyle\sum\nolimits_{k}}
E_{k}d_{\text{i}}p_{k}\geq0$, and $dS\equiv d_{\text{e}}S+d_{\text{i}}S$ with
$d_{\text{e}}S=-%
{\textstyle\sum\nolimits_{k}}
\ln p_{k}d_{\text{e}}p_{k}$ and $d_{\text{i}}S=-%
{\textstyle\sum\nolimits_{k}}
\ln p_{k}d_{\text{i}}p_{k}\geq0$ as a sum over microstates. One can easily
check that the microstate representations of these thermodynamic quantities
satisfy the thermodynamic identity \cite{Gujrati-Entropy2}
\begin{equation}
d_{\text{i}}Q=(T-T_{0})d_{\text{e}}S+Td_{\text{i}}S. \label{IrreversibleHeat}%
\end{equation}
The energy conservation in the first law can be applied to the exchange
process with the medium and the internal process within the system, separately
as follows: $d_{\text{e}}E=d_{\text{e}}Q-d_{\text{e}}W$ and $d_{\text{i}%
}E=d_{\text{i}}Q-d_{\text{i}}W$. As it is not possible to change the energy of
$\Sigma$ by internal processes, we conclude that $d_{\text{i}}E\equiv0$ so
that $d_{\text{i}}Q\equiv d_{\text{i}}W$ as noted above. This result will
guide us here for the simple model calculation for an isolated system (no
medium) for which $d_{\text{e}}p_{k}\equiv0$ so that $dp_{k}=d_{\text{e}}%
p_{k}$.

To demonstrate that the above definition of temperature, pressure, etc. is
consistent with the second law, we rewrite (\ref{IrreversibleHeat}) to express
$d_{\text{i}}S$ as a sum of two independent contributions%
\begin{equation}
d_{\text{i}}S=(1/T-1/T_{0})d_{\text{e}}Q+d_{\text{i}}Q/T.
\label{IrreversibleEntropy}%
\end{equation}
Both contributions must be nonnegative in accordance with the second law.
Thus, exchange heat $d_{\text{e}}Q$ always flows from hot to cold, and
$d_{\text{i}}W=d_{\text{i}}Q\geq0$. When $d_{\text{i}}W$ consists of several
independent contributions, each contribution must be nonnegative in accordance
with the second law. This proves our assertion.
\begin{figure}[ptb]%
\centering
\includegraphics[
trim=0.550526in 3.298699in 1.102753in 2.754425in,
height=2.0141in,
width=2.77in
]%
{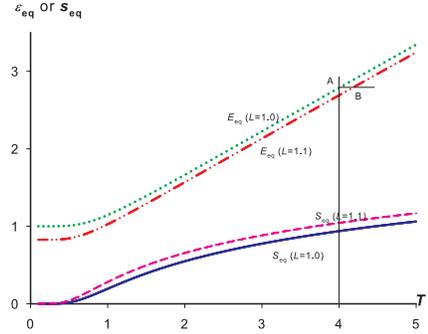}%
\caption{Equilibrium energy $\varepsilon_{\text{eq}}(T)$ (upper pair of
curves) and entropy $s_{\text{eq}}(T)\ $(lower pair of curves) for two
different box sizes $L=1.0$ and $1.1$\ obtained by using $p_{k,\text{eq}%
}(\beta,L)$. The point A on $L=1$\ corresponds to $T=4.0$ for which the energy
is $\varepsilon_{\text{eq}}\approx2.7859$. The point B on $L=1.1$ has the same
energy but has a higher temperature $T\approx4.1728$. }%
\label{Fig_Equilibrium_energy}%
\end{figure}

\textsc{Model }We consider a gas of $N$\ noninteracting identical
structureless spin-free nonrelativistic particles, each of mass $m$, confined
to a $1$-dimensional box with impenetrable walls and partitions, the latter
dividing the box into different sizes. The box is isolated so that
$d_{\text{e}}Q=0$. Initially, the gas is in thermodynamic \emph{equilibrium}
at temperature $T_{\text{i}}$ and pressure\ $P_{\text{i}}$ in state
A$_{\text{i,eq}}$, and is confined to a predetermined (such as the leftmost)
small part of the box of length $L_{\text{i}}$ by the leftmost partition. At
time $t=0$, the partition is instantaneously removed and the gas freely
expands to a box of size $L=\alpha L_{\text{i}}$, $\alpha>1$, imposed by the
next partition in a nonequilibrium fashion \cite{note2}. We wish to identify
the \emph{instantaneous} temperature and pressure of the gas as a function of
the box size $L$. \ 

Due to the lack of inter-particle interactions, we can focus on a single
particle, an extensively studied model in the literature but with a very
different emphasis \cite{Bender,Doescher,Stutz}. Here, we study it from the
current perspective. The particle only has non-degenerate eigenstates
(standing waves) whose energies are determined by $L$ and a quantum number
$k$; $p_{k}$\ denotes their probabilities. We use the energy scale
$\epsilon_{0}=\pi^{2}\hbar^{2}/2mL_{\text{i}}^{2}$ to measure the energy of
the eigenstate so that $\varepsilon_{k}(L)=k^{2}/\alpha^{2}$; the
corresponding eigenfunctions are given by%
\begin{equation}
\psi_{k}(x)=\left\langle x\right\vert \left.  k\right\rangle =\sqrt{2/L}%
\sin(k\pi x/L),\ k=1,2,\cdots. \label{eigenfunctions}%
\end{equation}
The pressure in the $k$th eigenstate is given by $P_{k}(L)\equiv
-\partial\varepsilon_{k}/\partial L=2\varepsilon_{k}(L)/L$ \cite{Landau-QM}.
The \emph{average} energy and entropy per particle, and the pressure are given
by (we suppress the $\left\{  p_{k}\right\}  $-dependence \emph{encoding} all
possible nonequilibrium states)
\begin{subequations}
\label{particle energy pressure}%
\begin{align}
\varepsilon(L)  &  \equiv%
{\textstyle\sum\nolimits_{k}}
p_{k}\varepsilon_{k},s(L)\equiv-%
{\textstyle\sum\nolimits_{k}}
p_{k}\ln p_{k}\label{particle energy}\\
P(L)  &  \equiv%
{\textstyle\sum\nolimits_{k}}
p_{k}P_{k}=2\varepsilon(L)/L. \label{particle pressure0}%
\end{align}

The \emph{equilibrium} state A$_{\text{eq}}(T,L)$ at \emph{dimensionless}
temperature $T$ (in the units of $\epsilon_{0}$)\ is given by the Boltzmann
law ($\beta\equiv1/T$) for $p_{k}$:%
\end{subequations}
\begin{equation}
p_{k\text{,eq}}(\beta,L)=\exp(-\beta\varepsilon_{k}(L))/Z_{0}(\beta,L);
\label{Equilibrium_pk}%
\end{equation}
$Z_{0}(\beta,L)\equiv%
{\textstyle\sum\nolimits_{k}}
\exp(-\beta\varepsilon_{k}(L))$ is the partition function. The equilibrium
macrostate is uniquely specified by $\{p_{k\text{,eq}}(\beta,L)\}$.

\textsc{Results }We plot $\varepsilon_{\text{eq}}(T,L)$ and $s_{\text{eq}%
}(T,L)$ in Fig. \ref{Fig_Equilibrium_energy} as a function of $T$ for two
different values of $L$; $P_{\text{eq}}=2\varepsilon_{\text{eq}}/L$. We
observe that $\varepsilon$ decreases as $L$ increases. To study expansion in
the isolated gas, for which $\varepsilon$ does not change \cite{Bender}, we
draw a horizontal line AB at $\varepsilon,$ which crosses the $L=1$ curve at
$T_{1\text{eq}}$, and the $L=1.1$ curve at $T_{2\text{eq}}$. For
$\varepsilon\simeq2.7859$ (see below)$,$ $T_{1\text{eq}}=4.0$, and
$T_{2\text{eq}}\simeq4.1728$. As the gas expands \emph{isoenergetically} from
$L_{\text{i}}=1.0$ to $L=1.1$, its temperature varies from $T_{1\text{eq}}$ to
eventually reach $T_{2\text{eq}}$ after the equilibration time $\tau
_{\text{eq}}$. However, we learn something more from the figure. If we
consider the temperature of the gas at some intermediate time $t$ during this
period, such as immediately after the \emph{free expansion} \cite{note2}, its
temperature $T(t)$ will continuously change towards $T_{2\text{eq}}$ in time.
The equilibrium entropy also increases with $L$ in an isothermal expansion, as
expected; see the vertical line through A at $T=4.0$.%
\begin{figure}[ptb]%
\centering
\includegraphics[
trim=0.550526in 1.238389in 0.826214in 4.679217in,
height=2.4751in,
width=3.4515in
]%
{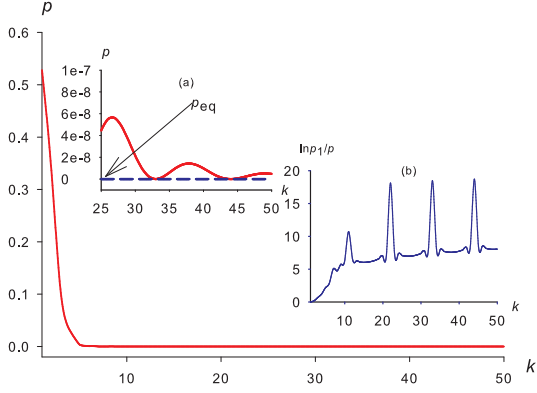}%
\caption{Eigenstate probabilities $p_{k}$ as a function of $k$ for $L=1.1$
after free expansion from state A$_{0}$. In the inset (a), we compare\ the
equilibrium probabilities $p_{k\text{,eq}}$ for $L=1.1$ ($T=4.18$) and $p_{k}$
in the main frame for higher $k$'s that clearly show oscillations. In the
inset (b), we plot $r\equiv\ln p_{1}/p_{k}$ that clearly shows oscillations
even for small $k$'s; these oscillation are not present in $p_{k\text{,eq}}$.
The curves in this figure are drawn for convenience. }%
\label{Fig-NonEqProb}%
\end{figure}

To identify $T(t)$, we proceed in three steps. In the first step, we
investigate the influence of \emph{quantum expansion} on the entropy $s$. The
gas is initially in a box of length $L_{\text{i}}$ with probabilities
$p_{k_{\text{i}}}$ of eigenstates $\left\vert k_{\text{i}}\right\rangle
\equiv\left\vert k,L_{\text{i}}\right\rangle $ and with energy and entropy per
particle $\varepsilon_{\text{i}}$, and $s_{\text{i}}$, respectively. For an
arbitrary state not in equilibrium or internal equilibrium, $p_{k}$ are
independent of the energies $\varepsilon_{k}$ of the $k$th microstate. We find
useful to deal with real probability "amplitude" $a_{k}$ determining $p_{k}$
($\equiv\left\vert a_{k}\right\vert ^{2}$) in the following. The gas directly
expands freely to a box of size $L_{1}$ or $L_{2}$, in each case starting from
$L_{\text{i}}$, and we calculate the amplitudes of various eigenstates
$\left\vert k_{\text{1}}\right\rangle \equiv\left\vert k,L_{\text{1}%
}\right\rangle $ and $\left\vert k_{\text{2}}\right\rangle \equiv\left\vert
k,L_{\text{2}}\right\rangle $\ in the two boxes:%
\[
a_{k_{1}}^{(\text{i})}=\sum_{k_{\text{i}}}a_{k_{\text{i}}}\left\langle
k_{1}\right\vert \left.  k_{\text{i}}\right\rangle ,~a_{k_{2}}^{(\text{i}%
)}=\sum_{k_{\text{i}}}a_{k_{\text{i}}}\left\langle k_{2}\right\vert \left.
k_{\text{i}}\right\rangle ,
\]
from which we calculate the entropies $s_{1}^{(\text{i})}$ and $s_{2}%
^{(\text{i})}$, respectively; the superscript is a \emph{reminder of the
memory effect} since these quantities depend on the initial state through
$p_{k_{\text{i}}}$. The coefficients $\left\langle k_{1}\right\vert \left.
k_{\text{i}}\right\rangle $, etc. are \cite{Bender}
\[
\left\langle k_{1}\right\vert \left.  k_{\text{i}}\right\rangle =\frac
{2k_{\text{i}}\alpha^{3/2}(-1)^{k_{\text{i}}}}{\pi(k_{1}^{2}-\alpha
^{2}k_{\text{i}}^{2})}\sin(\frac{k_{1}\pi}{\alpha}).
\]%
\begin{figure}[ptb]%
\centering
\includegraphics[
trim=0.824190in 3.430445in 0.551442in 2.476260in,
height=2.3229in,
width=3.2327in
]%
{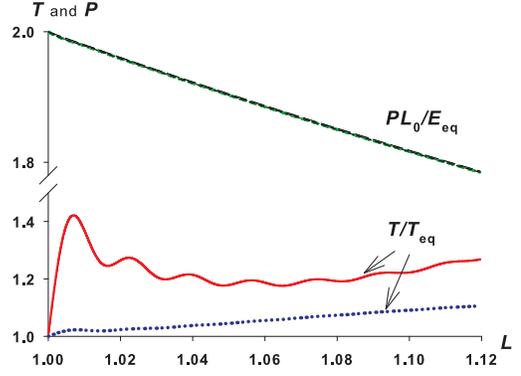}%
\caption{The normalized nonequilibrium temperature $T(L)/T_{\text{eq}%
}(L_{\text{i}})$ and pressure $P(L)L_{0}/\varepsilon_{\text{i}}(L_{\text{i}})$
for different $L$ after free expansion A$_{\text{i}}\rightarrow$A$(L)$. We
have taken $L_{\text{i}}=1.0$. The bottom two curves for $T(L)/T_{\text{eq}%
}(L_{0})$ correspond to $T_{\text{eq}}=1.0$ (solid) and $T_{\text{eq}}=4.0$
(dotted), respectively. The oscillations are more prominent at lower
equilibrium temperatures but there is an overall tendency to increase. The two
pressure curves for the two $T_{\text{eq}}$'s are almost indistinguishable on
the scale of the plot and can be used as the "exactness" of the computation.
Theoretically, the normalized pressure is independent of the temperature.}%
\label{Fig-NonEqT(L)-P(L)}%
\end{figure}

Because of the "deterministic" laws of quantum mechanics and the completeness
of the eigenstates, the amplitude $\sum_{k_{1}}a_{k_{1}}^{(\text{i}%
)}\left\langle k_{2}\right\vert \left.  k_{1}\right\rangle $ of the eigenstate
$\left\vert k_{2}\right\rangle $ after expansion from $L_{\text{i}}$ to
$L_{1}$\ to $L_{2}$\ is exactly $a_{k_{2}}^{(\text{i})}$. Thus, the entropy
$s_{2}^{(\text{i})}$ obtained from the direct expansion $L_{0}\rightarrow
L_{2}$ is the same as the entropy obtained from the expansion sequence
$L_{\text{i}}\rightarrow L_{1}\rightarrow L_{2}.$ We have also checked that
the two entropies are the same to within our numerical accuracy in our
computation. This means that the final ($L$)\ entropy has a memory of the
initial ($L_{\text{i}}$) state, but not of the paths from $L_{\text{i}}$ to
$L$. Thus, the entropy $s^{(\text{i})}(\varepsilon,L)$ in \emph{pure quantum
mechanical evolution} from a given \emph{initial} state is not a state
function of $\varepsilon$ and $L$. This is an important observation. \ 

The memory effect results in a nonequilibrium state. The consequences of the
latter can also be appreciated by considering the eigenstate probabilities
$p_{k}$ for different $k$, which is shown in the main frame in Fig.
\ref{Fig-NonEqProb} for $L=1.1$. It appears to fall off very rapidly, just as
$p_{k\text{,eq}}$. However, while $p_{k\text{,eq}}$ monotonically decreases
with $k$, $p_{k}$ has an \emph{oscillatory} behavior, as shown in the inset
(a) for $k$ between $25$ and $50$, where we compare the two probabilities;
here, the former is effectively zero. The fine structure of this oscillatory
behavior becomes obvious by considering the behavior of $\ln(p_{1}/p_{k})$,
which is plotted in the inset (b) for $k\geq1$. The oscillations are in
conformity with the presence of sine in $\left\langle k_{1}\right\vert \left.
k_{0}\right\rangle $, and should not be a surprise.

In the second step, we determine $T$ and $P$ for the nonequilibrium state
A$(\varepsilon_{\text{i}},L)$ in a box of size $L$ after free expansion from
$L_{\text{i}}$. The initial state A$_{\text{i}}$ is an equilibrium state
A$_{\text{i,eq}}(T=4.0$ $(\varepsilon_{\text{i}}\approx2.7859),L_{\text{i}%
}=1.0)$ for which $s_{\text{i,eq}}\approx0.94$. The entropy difference $\Delta
s\equiv s(L=1.1)-s_{\text{i,eq}}\approx1.06-0.94=0.12$ is \emph{positive},
which is expected in a free expansion. For the determination of the
temperature, we proceed as follows. We allow the gas to freely expand
($P_{0}^{\prime}=0$) from $L$ by a "differential" amount $dL\simeq0.0000001$
to $L^{\prime}$. In this differential expansion, $dQ=d_{\text{i}}Q\equiv
d_{\text{i}}W$ ($d_{\text{e}}Q=0$), and $d_{\text{i}}W=P(L)dL.$ We also
compute the change in the entropy $ds\equiv$ $s(L^{\prime})-s(L)$. The ratio
$PdL/ds$, see Eq. (\ref{dQ-dS}), determines the temperature $T$ of the
nonequilibrium gas. For $L=1.1$, we determine the temperature to be
$T(1.1)\approx4.365$ using this differential method, which lies outside the
equilibrium temperatures $T_{1\text{eq}}(1.0)=4.0$, and $T_{2\text{eq}%
}(1.1)=4.173$ quoted above. As we will show below, the higher nonequilibrium
temperature is due to "wider" microstate distribution relative to that for the
equilibrium state. The results for $T(L)$ for different $L$ in the free
expansion A$_{\text{i,eq}}\rightarrow$A$(\varepsilon_{\text{i}},L)$ are shown
in Fig. \ref{Fig-NonEqT(L)-P(L)}.

To add to the creditability of the above differential method for $T$, we apply
it to determine $T$ for the equilibrium state A$_{\text{i,eq}}(T=4.0,L=1.0)$.
For such a state, the ratio $r=\ln(p_{1}/p_{k})L^{2}/(k^{2}-1)$ is $r=1/T$ for
all $k$; see Eq. (\ref{Equilibrium_pk}). As $p_{k,\text{eq}}$ falls
exponentially with $k^{2}$, we truncate the number of microstates\ to $k\leq
k_{\text{tr}}$ for which $p_{k,\text{eq}}\geq10^{-15}$. This limits the number
of microstates to $k\leq k_{\text{tr}}=$ $13$. If we truncate using
$p_{k,\text{eq}}\geq10^{-22}$, then we need to consider $k\leq k_{\text{tr}%
}=15$. Thus, truncating the number of microstates to $k_{\text{tr}}$ is
computationally reasonable. The above calculation for the temperature with
$k\leq k_{\text{tr}}=$ $13$ gives $T=4.00000$ to the first five decimal
places, which adds to its creditability.%
\begin{figure}[ptb]%
\centering
\includegraphics[
trim=0.550526in 3.027664in 0.551377in 3.030969in,
height=2.5079in,
width=3.7317in
]%
{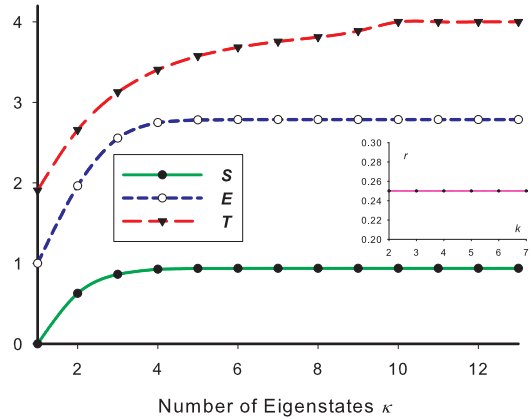}%
\caption{The effects of microstate numbers on the temperature, energy and
entropy after heat exchange at constant $L$. The initial state is
A$_{\text{i,eq}}$. The curves are guides to the eye. In the inset, we plot the
ratio $r\equiv\ln(p_{1}/p_{k})L^{2}/(k^{2}-1)$ for different microstates
indexed by $k$ for the choice $\kappa=7$ in the main figure. The ratio is
equal to the inverse temperature $1/4$ associated with A$_{\text{i,eq}}(4.0)$,
even though we have only seven microstates in the current state so that the
truncated state cannot be identified with an equilibrium state at $T=4.0$. }%
\label{Fig-Microstate_effects}%
\end{figure}

We now ask the following question:\ What will happen if we consider only the
first $\kappa$ microstates to determine the temperature, etc. by setting
$p_{k,\text{eq}}=0$ for $k>\kappa$. Such truncated states are obviously not
equilibrium states. To ensure that the probabilities add up to $1$, we
normalize the probabilities, which does not affect the ratio $r$, as follows:
$p_{k,\text{eq}}^{\prime}=p_{k,\text{eq}}/%
{\textstyle\sum}
p_{k,\text{eq}}$. The results for the temperature, energy and entropy are
shown in Fig. \ref{Fig-Microstate_effects}. In contrast, $r$ does not depend
on the value of $\kappa$ as shown in the inset for $\kappa\leq7$. But what we
observe is an interesting phenomenon. As the number $\kappa$ increases, that
is as the distribution gets "\emph{wider}," the temperature gets higher and
eventually gets to its limiting value of $4.0$.

The pressure is determined by Eq. (\ref{particle pressure0}) by setting
$\varepsilon(L)=\varepsilon_{\text{i}}$ so that $P(L)=2\varepsilon_{\text{i}%
}/L$. This is the statistical method (method 1) to compute $P(L)$.
Accordingly, $P(L)$ in an isoenergetic process is a decreasing function of
$L$. The ratio $P(L)L_{0}/\varepsilon_{\text{i}}$ is independent of the
$T_{\text{i}}$ of the initial state, which is confirmed by our computation as
shown by the upper curves for the two choices $T_{\text{i}}=1.0$ and
$T_{\text{i}}=4.0$ in Fig. \ref{Fig-NonEqT(L)-P(L)}. There is another way
(method 2) to determine the pressure in terms of the temperature, which is
based on a thermodynamic relation: $P/T=(\partial s/\partial L)_{\varepsilon}%
$. We use the ratio of the "differentials" $ds$\ and $dL$\ to determine $P/T$.
We now use the statistical temperature in Fig. \ref{Fig-NonEqT(L)-P(L)} in
this ratio to compute the thermodynamic pressure $P$. The results are found to
be indistinguishable from those shown in Fig. \ref{Fig-NonEqT(L)-P(L)} by
method 1, thus justifying our claim that the determination of our
nonequilibrium temperature is meaningful as the "internal" temperature of the
system in that the two different methods to determine the pressure give almost
identical values within our numerical accuracy.

As the memory of the initial state in $s_{1}^{(\text{i})},s_{2}^{(\text{i})},$
etc. cannot disappear by deterministic quantum evolution, some other mechanism
is required for equilibration to come about in which the nonequilibrium
entropy will gradually increase until it becomes equal to its equilibrium
value. One possible mechanism based on the idea of "chemical reaction" among
microstates has been proposed earlier \cite{Gujrati-Entropy1}. We will
consider the consequences of this approach elsewhere.


\begin{thebibliography}{99}
\bibitem {note-0}Following modern notation \cite{Prigogine,deGroot}, exchange
of any extensive quantity $Y(t)$ with the medium and its change within the
system carry the inner suffix e and i, respectively: $dY(t)\equiv d_{\text{e}%
}Y(t)+d_{\text{i}}Y(t)$. For example, we write $d_{\text{e}}S(t)$ and
$d_{\text{i}}S(t)$ as the two components of the entropy change $dS(t)$ in a
body: $dS(t)\equiv d_{\text{e}}S(t)+d_{\text{i}}S(t)$. A similar partition
$dp_{k}(t)=d_{\text{e}}p_{k}(t)+d_{\text{i}}p_{k}(t)$ can also be made.
\bibitem {note2}We neglect the time required for the eigenfunctions to expand
to the new length, which is much smaller than $\tau_{\text{eq}}$.
\end{thebibliography}
\end{document}